\documentclass[10pt,twocolumn,aps,pra,showpacs,superscriptaddress,floatfix,longbibliography,nofootinbib]{revtex4-2}

\usepackage{graphicx}
\usepackage{subfigure}
\usepackage[T1]{fontenc}
\usepackage{graphicx}
\usepackage[utf8]{inputenc}
\usepackage{amsmath, amsthm, amssymb,amsfonts}
\usepackage{color}
\usepackage{psfrag}
\usepackage{epsfig}
\usepackage{bbm}
\usepackage{bm}
\usepackage[colorlinks=true,citecolor=blue,urlcolor=magenta]{hyperref}
\usepackage[normalem]{ulem}
\usepackage{epstopdf}
\usepackage{graphicx}
\usepackage{stackengine,xcolor}
\usepackage{comment}

\definecolor{nred}{rgb}{0.9,0.1,0.1}
\definecolor{nblack}{rgb}{0,0,0}
\definecolor{nblue}{rgb}{0.2,0.2,0.8}
\definecolor{ngreen}{rgb}{0.2,0.6,0.2}

\usepackage{etoolbox}
\usepackage{bbold} % use \mathbb{1} to produce \openone
\newcommand{\beq}{\begin{eqnarray}}
\newcommand{\eeq}{\end{eqnarray}}

\newcommand{\rab}{{\varrho_{\text{AB}}}}
\newcommand{\rabin}{{\varrho_{\text{AB}}^{\rm in}}}
\newcommand{\rabout}{{\varrho_{\text{AB}}^{\rm out}}}

\newcommand{\Oab}{O_{\rm AB}}

\DeclareMathOperator{\tr}{tr}

\theoremstyle{definition}

\begin{document}

\title{Faithful certification of steering- and incompatibility-breaking channels}

\author{Po-Ting Hsu} 
\affiliation{Department of Physics, National Chung Hsing University, Taichung 40227, Taiwan}
\affiliation{TUM School of Natural Sciences, Technische Universit\"{a}t M\"{u}nchen, Garching 85748, Germany}

\author{Shin-Liang Chen}
\email{shin.liang.chen@email.nchu.edu.tw}
\affiliation{Department of Physics, National Chung Hsing University, Taichung 40227, Taiwan}
\affiliation{Physics Division, National Center for Theoretical Sciences, Taipei 106319, Taiwan}
\affiliation{Center for Quantum Frontiers of Research \& Technology (QFort), National Cheng Kung University, Tainan 701, Taiwan}

\date{ \today}

\begin{abstract}
We consider a quantum steering scenario where Alice prepares a bipartite state, sends one subsystem to Bob through a quantum channel, and aims to convince him that their shared state is entangled. Bob, however, will never be convinced if the channel is steering-breaking—that is, if the channel destroys steerability for any input state. To verify that a channel renders a state useless for demonstrating steering, one must consider all possible measurements performed by Alice and check if the conditional states Bob receives admit a local-hidden-state (LHS) model. This is generally a hard problem since it requires considering an infinite number of measurement combinations. Here, we show that the method proposed in [Phys. Rev. Lett. 117, 190401 (2016); Phys. Rev. Lett. 117, 190402 (2016)] can be utilized to tackle this problem. Furthermore, owing to the intimate relation between steering and measurement incompatibility, our approach can faithfully certify whether a channel destroys incompatibility for any set of measurements. Finally, we propose an experimental criterion for certifying steering- and incompatibility-breaking channels, which may be of independent interest.
\end{abstract}
\pacs{}

\maketitle
%%%%%%%%%%%%%%%%%%%%%%%%%%%%%%%%%%%%%%%%%%%%%%%%%%%%%%%%%%%%%%%%%%%%%%%%%

Quantum steering~\cite{Cavalcanti17,Uola2020Steering} is a phenomenon describing that one party (Alice) is able to remotely prepare a set of quantum states for the other distant party (Bob). Operationally, it can be defined as an entanglement certification task where the underlying state and one of the measurement devices are uncharacterized~\cite{Wiseman07}. That is, Alice tries to convince Bob that the state shared between them is entangled. From Bob’s perspective, he has no idea how the state was prepared and has no information about Alice’s measurement devices. Therefore, he can only ask Alice to perform some measurements and analyze the conditional states on his hand. If the set of conditional states, called the assemblage~\cite{Pusey13}, does not admit a local-hidden-state (LHS) model~\cite{Wiseman07}, hence demonstrating steering, then Bob is convinced that the shared state is entangled.

Closely related to steering is the concept of measurement incompatibility~\cite{Guhne2023}. In the quantum realm, there exist measurements that cannot be performed simultaneously, meaning they do not share a common joint measurement~\cite{Lahti03}. This property, known as measurement incompatibility, is not only a fundamental feature of quantum mechanics but also a necessary resource for demonstrating quantum steering~\cite{Quint14,Uola14}. Specifically, for a set of measurements to demonstrate steering on any quantum state, these measurements must be incompatible~\cite{Quint14,Uola14} (see also \cite{Uola15,Tavakoli2020Measurement,Porto2026Can_arXiv,Porto2026Measurement} for the intimate relation between steering and measurement incompatibility). Apart from steering, measurement incompatibility has been identified as a crucial resource for Bell nonlocality~\cite{Wolf09} and contextuality\cite{Xu19,Tavakoli2020Measurement}.

In practical scenarios, however, the transmission of quantum states or the implementation of measurements is inevitably subject to environmental noise, which is described by a quantum channel. If a channel is sufficiently noisy, it may completely destroy the useful resources. Such channels are referred to as steerability-breaking~\cite{HYKu2022} or incompatibility-breaking~\cite{Heinosaari2015} if they destroy the steerability or incompatibility for any input state or measurement assemblage, respectively (see \cite{Horodecki2003,Ruskai2003} for entanglement-breaking channels). Therefore, verifying whether a channel falls into these useless categories is crucial before performing further information tasks. Intuitively, this certification is a hard problem. To verify that a channel is steerability- or incompatibility-breaking, one has to check if the output states or measurements admit a classical model (LHS or jointly measurable model) for all possible input states~\cite{HYKu2022} and all possible measurements performed by Alice~\cite{Wiseman07,HYKu2022}. Here, we show that the method proposed in Refs.~\cite{Cavalcanti2016,Hirsch2016} and the results in Ref.~\cite{HYKu2022} can be effectively adapted to tackle this seemingly intractable problem. By explicitly considering a family of quantum channels, we derive conditions under which the channels destroy these quantum resources while potentially preserving entanglement. Finally, as the computation for certifying steering-breaking channels can be cast as a semidefinite program, we derive its dual program. We find that the dual program takes the form of a steering inequality. Consequently, the dual program can serve as an experimental criterion for certifying that the underlying channel is steering- and incompatibility-breaking.

\begin{figure}
\includegraphics[width=0.9\linewidth]{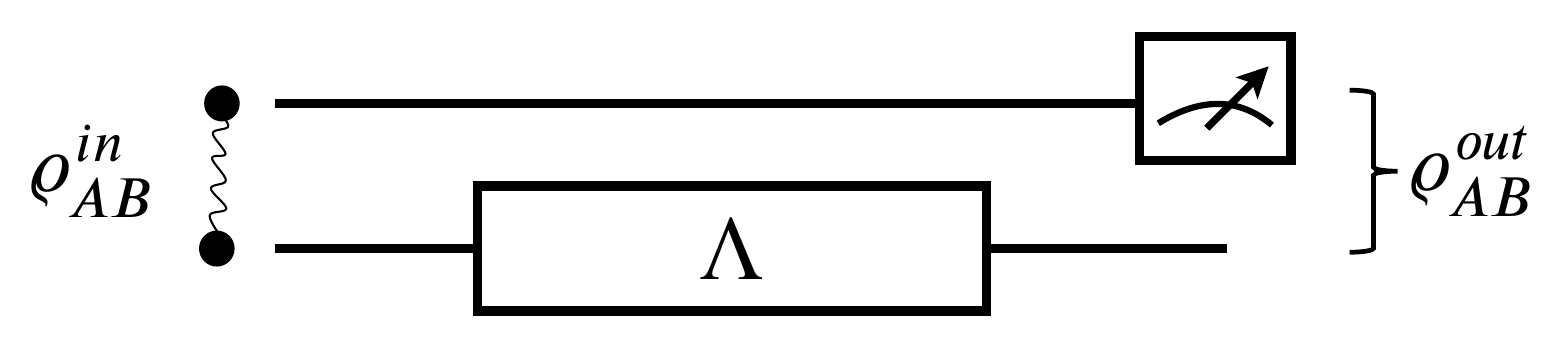}
\caption{In this work, we present an approach to verify whether a quantum channel $\Lambda$ is steering- and incompatibility-breaking for all input state $\rabin$.}
\label{Fig_channel}
\end{figure}

\textit{Quantum steering---}
Alice and Bob share a bipartite state $\rab$~\cite{Wiseman07,Cavalcanti17,Uola2020Steering}. Alice performs her $x$th measurement and obtains the outcome $a$. That measurement is a POVM $\{E_{a|x}\}_a$ with $E_{a|x}\succeq0$ and $\sum_a E_{a|x}=\openone$. Bob is left with the subnormalized states
\begin{equation}
\sigma_{a|x} = \tr_{\rm A}( E_{a|x}\otimes\openone \rab )\quad \forall a,x,
\label{Eq_Q_assemblage}
\end{equation}
which form the \emph{assemblage} $\{\sigma_{a|x}\}_{a,x}$~\cite{Pusey13}. The assemblage is \emph{unsteerable} when it admits a local-hidden-state (LHS) model~\cite{Wiseman07,Pusey13,SNC14},
\begin{equation}
\sigma_{a|x} = \sum_\lambda P(\lambda)P(a|x,\lambda)\sigma_{\lambda}\quad\forall a,x,
\label{Eq_LHS}
\end{equation}
in which Bob holds a fixed ensemble $\{\sigma_\lambda\}_\lambda$ and Alice merely heralds which member he has. Whether a given assemblage is unsteerable is decided by a semidefinite program~\cite{Pusey13,SNC14}.

A \emph{state} $\rab$ admits a LHS model when the assemblage it generates through Eq.~\eqref{Eq_Q_assemblage} is unsteerable for every number and every choice of measurements. Deciding this is far harder, since infinitely many measurements are involved. See Refs.~\cite{Werner89,Bowles2014,Jevtic2015,Bowles2015,Hirsch2017,ZhangYJ2024} for specific two-qubit families. References~\cite{Cavalcanti2016,Hirsch2016} showed that a finite set of measurements already suffices, and that construction is the tool we use below.

\textit{Constructing LHS models for quantum states---}
The method of Refs.~\cite{Cavalcanti2016,Hirsch2016} takes a finite set of measurements $\{M_{a|x}\}$ together with the continuous set $\mathcal{M}$ one wishes to certify. For a state $\xi_{\rm A}$ and any $M_a\in\mathcal{M}$ the shrunk measurement is
\begin{equation}
M_a^{r} := r M_a + (1-r)\tr(\xi_{\rm A}M_a)\openone.
\label{Eq_shrunk_meas}
\end{equation}
In Ref.~\cite{Hirsch2016}, the \emph{shrinking factor} $r$ is the largest value for which every $M_a^{r}$ is a convex combination of the elements of $\{M_{a|x}\}$~\cite{Hirsch2016}. In Ref.~\cite{Cavalcanti2016}, $r$ is the radius of the insphere of the polytope generated by the Bloch vectors of $\{M_{a|x}\}$ when taking $\mathcal{M}$ to be all projective qubit measurement.

With a fixed $r$, for a unit-trace operator $O_{\rm AB}$ (not necessarily positive semidefinite) and any quantum state $\xi_{\rm A}$, if
\begin{equation}
\{\tr_{\rm A}(M_{a|x}\otimes\openone~O_{\rm AB})\}_{a,x}~~\text{admits a LHS model},
\label{Eq_LHS_OAB}
\end{equation}
then
\begin{equation}
r O_{\rm AB} + (1-r) \xi_{\rm A}\otimes O_{\rm B}
\label{Eq_shrunk_state}
\end{equation}
also admits a LHS model for every measurement in $\mathcal{M}$. Here, $O_{\rm B}:=\tr_{\rm A} O_{\rm AB}$. Consequently, if the state $\rab$ can be decomposed as Eq.~\eqref{Eq_shrunk_state}, then $\rab$ admits a LHS model for all of $\mathcal{M}$. Determining whether it can be so decomposed is a semidefinite program (SDP):

\begin{equation}
\begin{aligned}
\text{given}~~&\rab, M_{a|x}, r\\
\text{find}~~&\Oab,\rho_\lambda\\
\text{s.t.}~~&\tr_{\rm A} (M_{a|x}\otimes \openone~ \Oab) = \sum_\lambda D(a|x,\lambda)\rho_{\lambda},~\forall a,x\\
&\rho_\lambda\succeq 0,\quad\forall \lambda\\
& \rab = r\Oab + (1-r)\frac{\openone}{d_{\rm A}}\otimes O_{\rm B}
\end{aligned}
\label{Eq_check_LHS}
\end{equation}
where $D(a|x,\lambda)$ are deterministic probabilities used for programming $P(a|x,\lambda)$ in Eq.~\eqref{Eq_LHS}. The first two constraints are from Eq.~\eqref{Eq_LHS_OAB}. The last constraint is from Eq.~\eqref{Eq_shrunk_state}, with $\xi_{\rm A}$ substituted by $\openone/d_{\rm A}$ to keep it linear. For a finite set $\{M_{a|x}\}$, Eq.~\eqref{Eq_check_LHS} is only a sufficient certification of a LHS model. A feasible test means that $\rab$ admits a LHS model for $\mathcal{M}$. An infeasible one leaves the steerability of $\rab$ inconclusive. When $\mathcal{M}$ is the set of all projective measurements, the test becomes necessary and sufficient as the number of measurements increases to infinity. Equivalently this is the limit $r\to1$. When $\mathcal{M}$ is the set of all POVMs, Refs.~\cite{Cavalcanti2016,Hirsch2016} take different approaches. Reference~\cite{Cavalcanti2016} starts from the projective result and then applies the depolarization map of Ref.~\cite{Acin2006}, which leaves a sufficient condition even in the limit $r\to1$. For both cases in \cite{Cavalcanti2016}, $r$ remains the insphere radius of the polyhedron formed by $\{M_{a|x}\}$. Reference~\cite{Hirsch2016} instead treats $r$ as the shrinking factor of $\mathcal{M}$ itself, so that Eq.~\eqref{Eq_check_LHS} remains unchanged and only the value of $r$ differs. In this case, the sequence of tests converges, therefore the certification is faithful for both projective measurements and POVMs.

\textit{Faithful certification of steering- and incompatibility-breaking channels---}
Now we turn to our main subject: Given a channel, how to verify if it destroys steerability for all input states, and how to verify if it destroys incompatibility for all input measurements?

The first question is equal to verifying if the output state admits a LHS model for all input states. First, we use Theorem 1 in Ref.~\cite{HYKu2022}, which states that a channel is steerability breaking if and only if it breaks steerability for the maximally entangled state. Therefore, it is sufficient to consider the maximally entangled state as the input state. Then, to check if the steerability of the output state is destroyed, we check if it admits a LHS model for any projective or POVM measurements. This can be resolved by the method of Ref.~\cite{Cavalcanti2016,Hirsch2016}, reviewed in the previous section.

To answer the second question, recall a channel $\Lambda$ is said to be incompatibility breaking for a class of measurement (e.g., POVM or projective measurements)~\cite{Heinosaari2015} if its dual $\Lambda^\dagger$ maps any POVM measurement to jointly measurable model~\cite{Quint14,Uola14}. That is, for all measurement assemblage $\{E_{a|x}\}_{a,x}$
\begin{equation}
\Lambda^\dagger(E_{a|x}) = \sum_\lambda P(a|x,\lambda)G_\lambda~\forall a,x.
\label{Eq_JM_model}
\end{equation}
The right-hand side of the above equation is the jointly measurable model~\cite{Quint14,Uola14}, implying that the statistics of different measurements $\{\{E_{a|x}\}_a\}_x$ can be simulated by a single measurement $\{G_\lambda\}_\lambda$. If a channel is incompatibility breaking, then any phenomena requiring incompatible measurements (e.g. nonlocality and steerability) cannot be demonstrated. In other words, such a channel is useless for tasks requiring incompatibility as a resource~\cite{Carmeli19,Skrzypczyk19,Uola19a,Designolle19}. 

To verify that a channel is incompatibility breaking, one has to check if all $E_{a|x}$ satisfy Eq.~\eqref{Eq_JM_model}. This seemingly intractable problem can be resolved by using Theorem 2 in Ref.~\cite{HYKu2022}, which states that a channel is steerability breaking if and only if it is incompatibility breaking. Therefore, once we find out a LHS model for the output state, we automatically verify that the channel breaks incompatibility for all POVM or projective measurements.

Finally, as the method of \cite{Cavalcanti2016,Hirsch2016} is faithful in principle, i.e., it can be used for certifying all non-steerable states, it is in principle faithful for certifying all steerability breaking and incompatibility breaking channels.

\textit{General procedures and an explicit example---}
We denote $\rabin$ and $\rabout$ as the input and output states, respectively. Namely,
\begin{equation}
\begin{aligned}
\rabout&= (\mathsf{id}\otimes\Lambda)(\rabin)\\
\end{aligned}
\label{Eq_Kraus}
\end{equation}
From the previous argument, $\rabin$ can be chosen as the maximally entangled state $|\Psi^+\rangle\langle\Psi^+|$, with $|\Psi^+\rangle:=(1/\sqrt{d})\sum_i|i\rangle\otimes|i\rangle$. Then we check if $\rabout$ admits a LHS model. To choose a set of projective measurements $M_{a|x}$, we use the property that every pair of vectors $\pm\hat{u}_x$ on the qubit Bloch sphere corresponds to a pair of projectors $\{M_{\pm|x}\}$ via the following relation:
\begin{equation}
M_{\pm|x} = \frac{1}{2}(\openone \pm \hat{u}_x\cdot \vec{\sigma}),
\end{equation}
where $\vec{\sigma}:=(X,Y,Z)$ is the set of Pauli matrices. Therefore, we choose vectors as vertices of various convex polyhedra. The meaning of $r$ in Eq.~\eqref{Eq_check_LHS} now is the radius of the insphere of each polyhedron. Having the information about $\rab$, $r$, and $M_{a|x}$, we can test the feasibility of SDP of Eq.~\eqref{Eq_check_LHS}.

\begin{figure}
\includegraphics[width=0.9\linewidth]{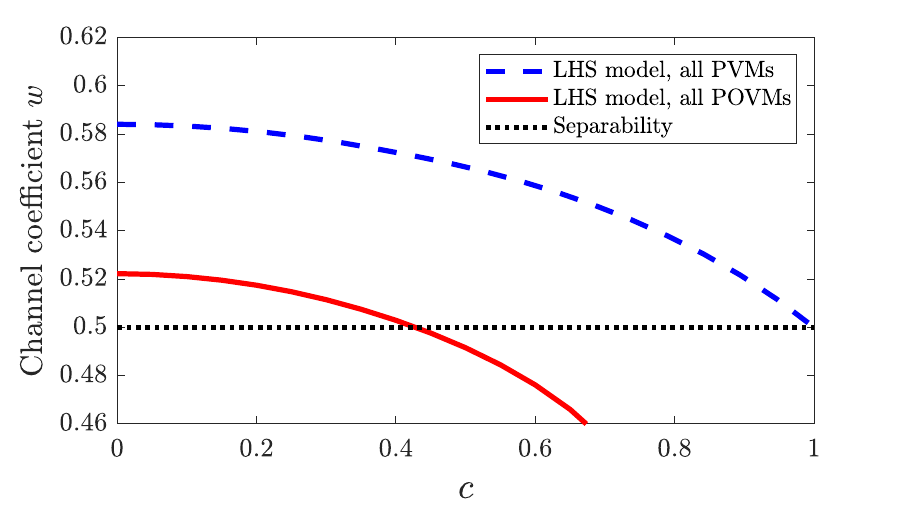}
\caption{The solution of Eq.~\eqref{Eq_check_LHS_max} for various values of $c$, which parametrizes the family of channels described by Eq.~\eqref{Eq_mixing_channel}. The black-dotted line is where the channel becomes entanglement breaking. Below the blue-dashed curve, the channel breaks steerability and incompatibility for all projective measurements. Below the red-solid curve, the channel breaks the two for all POVMs. Between either curve and the dotted line the entanglement survives. The two curves come from different measurement sets. For the result of projective measurements, we use the twelve measurements along the vertices of a rhombicuboctahedron~\cite{Cavalcanti2016}. For POVM, we use the six measurements along the vertices of an icosahedron, with $r=0.673$~\cite{Hirsch2016}.
}
\label{Fig_channels}
\end{figure}

The channels we consider in the following are noisy channels, linearly parametrized by a parameter $w\in [0,1]$, yielding the output state denoted by $\rabout(w)$. The effect of the channel increases as $w$ decreases. We are interested in the minimum effect of channel required to destroy the steerability of the input state. Therefore, we can make Eq.~\eqref{Eq_check_LHS} strictly feasible by maximizing $w$
\begin{equation}
\begin{aligned}
\text{given}~~&\rabout(w), M_{a|x}, r\\
\max_{w,\Oab,\rho_\lambda}~~& w\\
\text{s.t.}~~&\tr_{\rm A} (M_{a|x}\otimes \openone~ \Oab) = \sum_\lambda D(a|x,\lambda)\rho_{\lambda},~\forall a,x\\
&\rho_\lambda\succeq 0,\quad\forall \lambda\\
& \rabout(w) = r\Oab + (1-r)\xi_{\rm A}\otimes O_{\rm B}
\end{aligned}
\label{Eq_check_LHS_max}
\end{equation}

We consider several basic qubit channels, e.g., the bit-flip channel, the phase-flip (dephasing) channel, the decay channel, and the depolarizing channel. We find out that only the depolarizing channel provides a nontrivial result. That is, entanglement and steerability are destroyed at the same time for all the other channels.

For the depolarizing channel no new computation is needed. Acting on the two-qubit maximally entangled state, it outputs the isotropic state $\rabout(w)=v(w)|\Psi^+\rangle\langle\Psi^+|+(1-v(w))\openone/4$ with $v(w)=(4w-1)/3$ (see Appendix~\ref{Sec_App_optimal_F}). This state is entangled if and only if $v(w)>1/3$, that is $w>1/2$. Reference~\cite{Hirsch2016} constructed a LHS model covering all POVMs for the two-qubit Werner state at $v\simeq0.36$, which for two qubits is local unitarily equivalent to the isotropic state. Translating to the present parametrization gives
\begin{equation}
\frac{1}{2} < w \lesssim 0.52.
\label{Eq_depol_window}
\end{equation}
In this range the depolarizing channel breaks steerability for all POVMs, and hence breaks incompatibility by Theorem 2 of Ref.~\cite{HYKu2022}. Since the output state is still entangled, this is an explicit channel which is incompatibility breaking but not entanglement breaking.

To demonstrate more cases where the steerability is destroyed while the entanglement is preserved, we artificially create a family of channels by mixing the bit-flip channel and the depolarizing channel as follows:
\begin{equation}
\Lambda(\rho) = c\Lambda^{\rm b.f.}(\rho) + (1-c)\Lambda^{\rm depol.}(\rho),
\label{Eq_mixing_channel}
\end{equation}
where $\Lambda^{\rm b.f.}(\cdot) = w\rho + (1-w)X\rho X$ and $\Lambda^{\rm depol.}(\cdot) = w\rho + ((1-w)/3)(X\rho X + Y\rho Y + Z\rho Z)$. We consider the range $0\leq c\leq 1$ and run SDP~\eqref{Eq_check_LHS_max} with $12$ measurements. The result is the dashed curve of Fig.~\ref{Fig_channels}. The region above the dotted line and below that curve is where the steerability is destroyed for all projective measurements while the entanglement is preserved. The code for implementing the computation can be found in \cite{PTHsu2026_code}.

The POVM case depends on which route is taken. With the method of Ref.~\cite{Cavalcanti2016}, we cannot certify that the channel is steering breaking for any value of $r$ (see Appendix~\ref{Sec_App_Cavalcanti}). We therefore turn to the construction of Ref.~\cite{Hirsch2016}, whose shrinking factor is defined directly for the set of POVMs. Taking the six projective measurements along the vertices of the icosahedron, the shrinking factor of the two-qubit POVM set is $r=0.673$~\cite{Hirsch2016}. Using that value in Eq.~\eqref{Eq_check_LHS_max} and scanning over $c$ gives the solid curve of Fig.~\ref{Fig_channels}. At $c=0$ we obtain $w=0.5222$, corresponding to $v=0.363$. This reproduces the value quoted in Ref.~\cite{Hirsch2016} for the two-qubit Werner state and recovers Eq.~\eqref{Eq_depol_window}. The curve stays above the entanglement-breaking line for $c\lesssim0.42$. Throughout that range the channel breaks steerability for all POVMs while preserving entanglement, and hence breaks incompatibility by Theorem 2 of Ref.~\cite{HYKu2022}. The curve uses six measurements rather than the twelve behind the projective one, since we take the POVM shrinking factor directly from Ref.~\cite{Hirsch2016}.

\textit{Experimental criteria for certifying steerability and incompatibility breaking channels---} Using Eq.~\eqref{Eq_check_LHS_max} (or equivalently, Eq.~\eqref{Eq_check_LHS}) to verify steerability and incompatibility breaking channels, we have to know the exact form of the state $\rabout$. Experimentally, this means reconstructing the joint state by full tomography, which becomes costly as the dimension grows. In what follows, we propose a certification that avoids the state tomography. It requires only the expectation values of Bob's local observables on the conditional states. To do so, recall that every SDP has a dual form. The dual of Eq.~\eqref{Eq_check_LHS} takes the following form (see Appendix \ref{Sec_App_dual} for the derivation):
\begin{equation}
    \begin{aligned}
        \min ~~&\text{tr}\sum_{a,x} F_{a|x}\rho_{a|x}\\
        \text{s.t.}~~& \sum_{a,x} D(a|x,\lambda) F_{a|x}\succeq 0\\
        & \text{tr} \sum_{a,x,\lambda} D(a|x,\lambda) F_{a|x} = 1
    \end{aligned}
    \label{Eq_dual}
\end{equation}
where
\begin{equation}
    \rho_{a|x}:=\frac{1}{r}\sigma_{a|x} + \frac{r-1}{r}\sigma_{a|x}^{\rm m}.
    \label{Eq_affine}
\end{equation}
Here, $\sigma_{a|x}:=\tr_{\rm A}(M_{a|x}\otimes\openone~\rab)$ is the assemblage created by performing measurement $M_{a|x}$ on the state $\rab$, and
\begin{equation}
    \sigma_{a|x}^{\rm m}:=\tr_{\rm A}[(M_{a|x}\otimes\openone)(\xi_{\rm A}\otimes\tr_{\rm A}\rab)] = P_\xi(a|x)\cdot\tr_{\rm A}\rab,
\end{equation}
with $P_\xi(a|x):=\tr(M_{a|x}\xi_{\rm A})$, is the unsteerable assemblage created by performing $M_{a|x}$ on the separable state $\xi_{\rm A}\otimes\tr_{\rm A}\rab$.

The form of Eq.~\eqref{Eq_dual} is exactly a steering inequality in a standard steering scenario~\cite{Cavalcanti17}. The first constraint makes its value non-negative whenever $\rho_{a|x}$ admits a LHS model, so a negative solution certifies steerability of $\rho_{a|x}$. Consequently, while one concludes nothing about the state if a standard steering inequality holds, a non-negative solution of Eq.~\eqref{Eq_dual} certifies a LHS model of $\rab$.

A brief procedure for a two-qubit experiment is as follows. One defines a target state $\rabout$, a set of measurements $M_{a|x}$, and the state $\xi_{\rm A}\otimes\tr_{\rm A}(\rabout)$. This fixes $\rho_{a|x}$, and Eq.~\eqref{Eq_dual} yields the associated optimal steering inequality $F_{a|x}$. Decomposing each $F_{a|x}$ in the Pauli basis turns the objective function of Eq.~\eqref{Eq_dual} into a function of the probabilities $P(a|x)$ and Bob's local expectation values alone (see Appendix~\ref{Sec_App_Pauli}). If its value on the observed data is non-negative, we can certify that the channel is steerability and incompatibility breaking. Otherwise no conclusion is drawn. Note that a steering inequality obtained here as the optimal solution for one state cannot in general be reused to certify a different state since a steerable state may still return a non-negative value on that inequality. There are nevertheless families along which the optimal $F_{a|x}$ does not change. The depolarizing channel is one instance, so a single steering inequality verifies it across the whole family (see Appendix~\ref{Sec_App_optimal_F}).

\textit{Conclusion and discussion---}
In this work, we present an approach to verify whether a quantum channel is steering- and incompatibility-breaking. Specifically, we apply the algorithmic construction of local-hidden-state (LHS) models proposed in Refs.~\cite{Cavalcanti2016,Hirsch2016} and combine it with the theorems established in Ref.~\cite{HYKu2022} to show that evaluating the channel's action on a maximally entangled state is sufficient for this certification task. Furthermore, as this computation can be cast as a SDP, we derive its dual program, which takes the form of a steering inequality. This dual formulation serves as an experimentally friendly criterion that bypasses the need for full quantum state tomography.

Importantly, we highlight a fundamental limitation of the steering inequality. An inequality optimized for one particular state cannot, in general, be reused to faithfully certify the LHS model of a different state. Satisfying such a suboptimal inequality does not rule out the steerability of the target state. However, we show that the family of depolarizing channels is an exception where the optimal steering inequality remains invariant. This invariance allows a single steering inequality to verify whether an underlying depolarizing channel is steering- and incompatibility-breaking. A potential direction for future work is to investigate whether this invariance holds for other noisy channels or higher-dimensional systems, which could further simplify the experimental certification of quantum channels.

\textit{Note added---} While completing this manuscript, we became aware of a recent work \cite{YJZhang2026Exact} deriving the exact incompatibility-breaking criterion for unital qubit channels with a different approach.

\acknowledgements S.-L.~C. acknowledges the support of the National Science and Technology Council (NSTC) Taiwan (Grant No. NSTC 114-2628-M-005-001- and 115-2628-M-005-001-) and National Center for Theoretical Sciences Taiwan (Grant No. NSTC 115-2124-M-002-014-). P.-T.~H. acknowledges the support of the National Science and Technology Council (NSTC) Taiwan (Grant No. 113-2813-C-005-009-M) and Center for Quantum Frontiers of Research \& Technology (QFort), National Cheng Kung University, Tainan, Taiwan.

\clearpage
\onecolumngrid
\appendix

\section{The output states for the depolarizing channel have the same optimal $F_{a|x}$}\label{Sec_App_optimal_F}
In this section, we show that the output state $\rabout$ for the depolarizing channel have the same optimal steering inequality $F_{a|x}$. First, let us write down the dual SDP again (i.e., Eq.~\eqref{Eq_dual} in the main text):
\begin{equation}
    \begin{aligned}
        \min ~~&\text{tr}\sum_{a,x} F_{a|x}\rho_{a|x}\\
        \text{s.t.}~~& \sum_{a,x} D(a|x,\lambda) F_{a|x}\succeq 0\\
        & \text{tr} \sum_{a,x,\lambda} D(a|x,\lambda) F_{a|x} = 1
    \end{aligned}
    \label{EqApp_dual}
\end{equation}
where
\begin{equation}
    \rho_{a|x}:=\frac{1}{r}\sigma_{a|x} + \frac{r-1}{r}\sigma_{a|x}^{\rm m},
    \label{EqApp_affine}
\end{equation}
with
\begin{equation}
    \sigma_{a|x}=\tr_{\rm A}(M_{a|x}\otimes\openone~\rabout)\quad\quad\text{and}\quad\quad
    \sigma_{a|x}^{\rm m}=P_\xi(a|x)\cdot\tr_{\rm A}(\rabout).
\end{equation}
The first step is to compute the output state $\rabout$ for the depolarizing channel, which is defined as
\begin{equation}
    \Lambda^{\rm depol}(\rho) = w\rho + \frac{1-w}{3}(X\rho X + Y\rho Y + Z\rho Z),
\end{equation}
where $0\leq w\leq 1$. When the channel acts on one half of the maximally entangled state $\vert{}\Psi^+\rangle$, the output state is
\begin{equation}
    \begin{aligned}
        \rabout &= (\mathsf{id} \otimes \Lambda)(\vert{}\Psi^+\rangle\langle\Psi^+\vert{})\\
        & =w \vert{}\Psi^+\rangle\langle\Psi^+\vert{} + \frac{1-w}{3} \sum_{i} (\openone \otimes \sigma_i) \vert{}\Psi^+\rangle\langle\Psi^+\vert{} (\openone \otimes \sigma_i)
    \end{aligned}
\end{equation}
where $\{\sigma_i\}:=\{X,Y,Z\}$. In a two-qubit system, the four Bell states form a complete orthonormal basis, and their equal mixture yields the completely mixed state $\openone/4$. Consequently, the Pauli noise terms can be rewritten as:
\begin{equation}
    \sum_{i} (\openone \otimes \sigma_i) \vert{}\Psi^+\rangle\langle\Psi^+\vert{} (\openone \otimes \sigma_i) = 4\left(\frac{\openone}{4}\right) - \vert{}\Psi^+\rangle\langle\Psi^+\vert{}
\end{equation}
Substituting this back into the output state yields:
\begin{equation}
    \begin{aligned}
        \rabout(w) &= \left( w - \frac{1-w}{3} \right) \vert{}\Psi^+\rangle\langle\Psi^+\vert{} + \frac{4(1-w)}{3} \left( \frac{\openone}{4} \right)\\
        &=v(w) \vert{}\Psi^+\rangle\langle\Psi^+\vert{} + (1-v(w)) \frac{\openone}{4}
    \end{aligned}
\end{equation}
where we define $v(w):=\frac{4w-1}{3}$.

Next, we compute $\sigma_{\rm m}$ and $\sigma_{a|x}$:
\begin{equation}
    \sigma_{a\vert{}x}^m = P_\xi(a\vert{}x) \rho_B = \tr\left(M_{a\vert{}x} \frac{\openone}{2}\right) \frac{\openone}{2} = \frac{1}{2} \cdot \frac{\openone}{2} = \frac{\openone}{4}
\end{equation}
The second and third equalities are obtained, respectively, by choosing $\xi_{\rm A} = \openone/2$ and by the fact that $M_{a|x}$ are all projective. Note that $\sigma_{a|x}^{\rm m}$ is independent of $w$. For $\sigma_{a|x}$, we have
\begin{equation}
    \begin{aligned}
        \sigma_{a|x} = v(w) \sigma_{a\vert{}x}^{\Psi^+} + (1-v(w))\frac{\openone}{4}
    \end{aligned}
\end{equation}
where $\sigma_{a\vert{}x}^{\Psi^+}:=\tr_{\rm A}(M_{a|x}\otimes\openone|\Psi^+\rangle\langle\Psi^+|)$ is the assemblage generated from the maximally entangled state.
Substituting $\sigma_{a\vert{}x}(w)$ and the constant $\sigma_{a\vert{}x}^m$ into the affine combination defined in Eq.~\eqref{EqApp_affine}, we obtain:
\begin{equation}
    \rho_{a\vert{}x}(w) = \frac{1}{r} \sigma_{a\vert{}x}(w) + \frac{r-1}{r} \sigma_{a\vert{}x}^m
    = \frac{v(w)}{r} \sigma_{a\vert{}x}^{\Psi^+} + \frac{r-v(w)}{r} \frac{\openone}{4}
\end{equation}

We now compute minimizer (i.e., the optimal $F_{a|x}$) of Eq.~\eqref{EqApp_dual}. Defining $I_s(F, w)$ as the objective function, we have
\begin{equation}
    \begin{aligned}
        F_{a|x}^{\rm opt}:=\min_F I_s(F, w) &= \min_F \tr\sum_{a,x} F_{a\vert{}x} \rho_{a\vert{}x}(w)\\
        &= \min_F \left[ \frac{v(w)}{r} \underbrace{\tr\sum_{a,x} F_{a\vert{}x} \sigma_{a\vert{}x}^{\Psi^+}}_{:= C_1(F)} + \frac{r-v(w)}{r} \underbrace{\tr\sum_{a,x} F_{a\vert{}x} \frac{\openone}{4}}_{:= C_2(F)} \right]
    \end{aligned}
\end{equation}
where $F:=\{F_{a|x}\}_{a,x}$. Note that the second constraint of Eq.~\eqref{EqApp_dual} imposes the constraint $tr\sum_{a,x,\lambda} D(a\vert{}x,\lambda)F_{a\vert{}x} = 1$. Because $D(a\vert{}x,\lambda)$ represents deterministic probability distributions, summing over all possible hidden-variable strategies $\lambda$ yields a uniform constant factor for each measurement outcome. This strictly restricts $tr\sum_{a,x} F_{a\vert{}x}$ to be a constant, regardless of the specific choice of the feasible variables $F_{a\vert{}x}$. Consequently, $C_2(F)$ is a constant, independent of $F$. The minimizer $F_{a|x}^{\rm opt}$ can then be written as
\begin{equation}
    \begin{aligned}
        &F_{a|x}^{\rm opt}=\\
        &\arg\min_F I_s(F, w) = \arg\min_F \left( \frac{v(w)}{r} C_1(F) + \frac{r-v(w)}{r} C_2(F)  \right) = \arg\min_F ~C_1(F) = \arg\min_F \tr\sum_{a,x} F_{a\vert{}x} \sigma_{a\vert{}x}^{\Psi^+}
    \end{aligned}
\end{equation}
where the third equality is due to the fact that adding a constant or multiplying by a positive scalar does not affect the optimizer of an optimization problem. The scalar here is $v(w)/r$, so the step requires $v(w)>0$, that is $w>1/4$. This demonstrates that the optimal steering inequality $F_{a\vert{}x}^{\rm opt}$ is completely determined by the assemblage $\sigma_{a\vert{}x}^{\Psi^+}$ and remains invariant across the family of states generated by the depolarizing channel for $w>1/4$. An isotropic state with $v(w)\le0$ is separable, so the excluded range contains no steerable states.

\section{The POVM criterion of Ref.~\cite{Cavalcanti2016} applied to the depolarizing channel}\label{Sec_App_Cavalcanti}
As shown in the main text, given the input state as the maximally entangled two-qubit state, the output states of the depolarizing channel are the two-qubit isotropic states:
\begin{equation}
    \rabout = v\,|\Psi^+\rangle\langle\Psi^+| + (1-v)\frac{\openone}{4} =: \rho_{\rm iso}(v),
\end{equation}
with $v=v(w)=(4w-1)/3$ as in the main text. We would like to prove the following observation.

\textbf{Observation 1} Given $\rabout=\rho_{\rm iso}(v)$, the solution of the following SDP~\cite{Cavalcanti2016} is upper bounded by $1/4$:
\begin{equation}
    \begin{aligned}
        \text{given}~~& M_{a|x}, r, \gamma_{\rm A}\\
        \max_{v,\Oab,\rho_\lambda}~~&v\\
        \text{s.t.}~~& \tr_{\rm A} (M_{a|x}\otimes \openone~ \Oab) = \sum_\lambda D(a|x,\lambda)\rho_{\lambda},~\forall a,x\\
&\rho_\lambda\succeq 0,\quad\forall \lambda\\
& \rabout = \frac{1}{d_{\rm A}}\Big[r\Oab + (1-r)\frac{\openone_{\rm A}}{d_{\rm A}}\otimes O_{\rm B}\Big] + \frac{d_{\rm A}-1}{d_{\rm A}}\,\gamma_{\rm A}\otimes O_{\rm B},
    \end{aligned}
    \label{EqApp_LHS_POVM}
\end{equation}
with $d_{\rm A}=2$.

\begin{proof}
First, from the last constraint of the above SDP, we can see that $\tr_{\rm A} \rabout = O_{\rm B}$. Together with the fact that $\tr_{\rm A} \rho_{\rm iso}(v) = \openone / 2$, we obtain
\begin{equation}
    O_{\rm B} = \frac{\openone}{2}.
\end{equation}
Then, we define the term in the bracket of the same constraint as
    \begin{equation}
    \begin{aligned}
        \tau_{\rm AB}:=&\;r\Oab + (1-r)\frac{\openone_{\rm A}}{2}\otimes O_{\rm B}\\
        =&\;2\rabout - \gamma_{\rm A}\otimes \frac{\openone}{2},
    \end{aligned}
    \end{equation}
where the second line solves the last constraint of Eq.~\eqref{EqApp_LHS_POVM} for the bracket and uses $O_{\rm B}=\openone/2$. Applying the twirl $\mathcal{T}(X):=\int dU\,(U\otimes U^*)\,X\,(U\otimes U^*)^\dagger$, with $dU$ the Haar measure, on the above equation, we obtain
\begin{equation}
    \begin{aligned}
        \mathcal{T}\big(\tau_{\rm AB}\big) &= \mathcal{T}\big(2\rabout\big) - \mathcal{T}\big(\gamma_{\rm A}\otimes\frac{\openone}{2}\big)\\
        &=2\rho_{\rm iso}(v) - \frac{\openone}{4}\\
        &=\rho_{\rm iso}(2v).
    \end{aligned}
\end{equation}
From the first two constraints of Eq.~\eqref{EqApp_LHS_POVM}, the assemblage of $O_{\rm AB}$ with respect to the finite set $\{M_{a|x}\}$ admits a LHS decomposition. Theorem 1 of Ref.~\cite{Cavalcanti2016} then guarantees that $\tau_{\rm AB}$ admits a LHS model for all projective measurements. The twirl $\mathcal{T}$, being a mixture of local unitaries, does not create steerability, so $\mathcal{T}(\tau_{\rm AB})$ admits such a model as well. The two-qubit isotropic state is unsteerable under all projective measurements only up to visibility $1/2$~\cite{Werner89,Wiseman07}, so $2v\leq 1/2$. Therefore, we have
\begin{equation}
    v\leq \frac{1}{4}.
\end{equation}
\end{proof}

The bound $v=1/4$ is equal to $w=7/16$ in the parametrization used in the main text. The bound holds for every $r$ and every $\gamma_{\rm A}$. An isotropic state is entangled only for $v>1/3$. Therefore, this criterion certifies nothing outside the separable set. Running it with $12$ measurements gives $v\approx0.223$, close to the bound.

\section{Derivation of the dual of Eq.~\eqref{Eq_check_LHS}}\label{Sec_App_dual}
In this appendix, we explicitly derive the dual semidefinite program (SDP) presented in Eq.~\eqref{Eq_dual} from the primal SDP in Eq.~\eqref{Eq_check_LHS}. First we rewrite the feasibility problem Eq.~\eqref{Eq_check_LHS} to a strictly feasible problem:
\begin{equation}
\begin{aligned}
    \max \quad & \mu \\
    \text{s.t.} \quad & \text{tr}_{\rm A}[M_{a|x} \otimes \openone_{\rm B} O_{\rm AB}] = \sum_{\lambda} D(a|x,\lambda) \rho_{\lambda} \quad \forall a,x, \\
    & \rho_{\lambda} \succeq \mu \cdot \openone, \\
    & \rho_{\rm AB} = r O_{\rm AB} + (1-r) \xi_{\rm A} \otimes \text{tr}_{\rm A}(O_{\rm AB}),
\end{aligned}
\end{equation}
where $\xi_{\rm A} = \openone/d_{\rm A}$. 

We introduce the Hermitian dual variables $F_{a|x}$ for the first equality constraint, $Y_{\lambda} \succeq 0$ for the inequality constraint, and $Z$ for the last constraint. The Lagrangian $\mathcal{L}$ is constructed as follows:
\begin{equation}
\begin{aligned}
    \mathcal{L} &= \mu + \sum_{a,x} \text{tr} \left\{ F_{a|x} \left[ \text{tr}_{\rm A}(M_{a|x} \otimes \openone_{\rm B} O_{\rm AB}) - \sum_{\lambda} D(a|x,\lambda) \rho_{\lambda} \right] \right\} \\
    &\quad + \sum_{\lambda} \text{tr} \left[ Y_{\lambda} (\rho_{\lambda} - \mu \cdot \openone) \right] \\
    &\quad + \text{tr} \left[ Z \left( \rho_{\rm AB} - r O_{\rm AB} - (1-r)\xi_{\rm A} \otimes \text{tr}_{\rm A}(O_{\rm AB}) \right) \right].
\end{aligned}
\end{equation}

By rearranging the terms with respect to the primal variables $\rho_{\lambda}$, $\mu$, and $O_{\rm AB}$, we obtain:
\begin{equation}
\begin{aligned}
    \mathcal{L} &= \text{tr} \sum_{\lambda} \left[ \left( -\sum_{a,x} D(a|x,\lambda) F_{a|x} + Y_{\lambda} \right) \rho_{\lambda} \right] + \left( 1 - \text{tr} \sum_{\lambda} Y_{\lambda} \right) \mu \\
    &\quad + \text{tr} \left\{ \left[ \sum_{a,x} M_{a|x} \otimes F_{a|x} - r Z - (1-r) \openone_{\rm A} \otimes \text{tr}_{\rm A}(Z (\xi_{\rm A} \otimes \openone_{\rm B})) \right] O_{\rm AB} \right\} \\
    &\quad + \text{tr}(Z \rho_{\rm AB}).
\end{aligned}
\end{equation}

For the dual problem to be bounded, the Lagrangian must be independent of the unconstrained primal variables $\rho_{\lambda}$, $\mu$, and $O_{\rm AB}$, which imposes the following dual constraints:
\begin{align}
    -\sum_{a,x} D(a|x,\lambda) F_{a|x} + Y_{\lambda} &= 0, \label{eq:dual_c1} \\
    1 - \text{tr} \sum_{\lambda} Y_{\lambda} &= 0, \label{eq:dual_c2} \\
    \sum_{a,x} M_{a|x} \otimes F_{a|x} - r Z - (1-r) \openone_{\rm A} \otimes \text{tr}_{\rm A}(Z (\xi_{\rm A} \otimes \openone_{\rm B})) &= 0, \label{eq:dual_c3} \\
    Y_{\lambda} &\succeq 0. \label{eq:dual_c4}
\end{align}

From Eq.~\eqref{eq:dual_c1} and Eq.~\eqref{eq:dual_c4}, we have:
\begin{equation}
    Y_{\lambda} = \sum_{a,x} D(a|x,\lambda) F_{a|x} \succeq 0.
\end{equation}
Substituting this into Eq.~\eqref{eq:dual_c2} yields the trace constraint:
\begin{equation}
    \text{tr} \sum_{a,x,\lambda} D(a|x,\lambda) F_{a|x} = 1.
\end{equation}

To simplify the objective function $\min \text{tr}(Z \rho_{\rm AB})$, we solve for $Z$ using Eq.~\eqref{eq:dual_c3}:
\begin{equation}
    Z + \frac{1-r}{r} \left[ \openone_{\rm A} \otimes \text{tr}_{\rm A}(Z (\xi_{\rm A} \otimes \openone_{\rm B})) \right] = \frac{1}{r} \sum_{a,x} M_{a|x} \otimes F_{a|x}.
\end{equation}
By defining $s := -\frac{1-r}{r(r + (1-r)\text{tr}\xi_{\rm A})}$, and noting that $\text{tr}\xi_{\rm A} = 1$, we find $s = \frac{r-1}{r}$. The general solution for $Z$ takes the form:
\begin{equation}
    Z = \frac{1}{r} \sum_{a,x} M_{a|x} \otimes F_{a|x} + s \cdot \openone_{\rm A} \otimes \text{tr}_{\rm A} \left[ \left(\sum_{a,x} M_{a|x} \otimes F_{a|x}\right) (\xi_{\rm A} \otimes \openone_{\rm B}) \right].
\end{equation}

We can now substitute $Z$ back into the dual objective function $\text{tr}(Z \rho_{\rm AB})$. Let us split this into two parts: $\text{tr}(Z \rho_{\rm AB}) = \text{\textcircled{1}} + \text{\textcircled{2}}$.
The first part is:
\begin{equation}
\begin{aligned}
    \text{\textcircled{1}} &= \text{tr} \left[ \left( \frac{1}{r} \sum_{a,x} M_{a|x} \otimes F_{a|x} \right) \rho_{\rm AB} \right] \\
    &= \frac{1}{r} \sum_{a,x} \text{tr}_{\rm B} \left[ F_{a|x} \text{tr}_{\rm A}(M_{a|x} \otimes \openone_{\rm B} \rho_{\rm AB}) \right] \\
    &= \frac{1}{r} \text{tr} \sum_{a,x} F_{a|x} \sigma_{a|x},
\end{aligned}
\end{equation}
where $\sigma_{a|x} := \text{tr}_{\rm A}(M_{a|x} \otimes \openone_{\rm B} \rho_{\rm AB})$.

The second part is derived by cyclically permuting the partial traces:
\begin{equation}
\begin{aligned}
    \text{\textcircled{2}} &= s \cdot \text{tr} \left\{ \left[ \openone_{\rm A} \otimes \text{tr}_{\rm A} \left( \sum_{a,x} (M_{a|x} \otimes F_{a|x})(\xi_{\rm A} \otimes \openone_{\rm B}) \right) \right] \rho_{\rm AB} \right\} \\
    &= s \cdot \sum_{a,x} \text{tr}_{\rm B} \left\{ \text{tr}_{\rm A} \left[ (M_{a|x} \otimes F_{a|x})(\xi_{\rm A} \otimes \openone_{\rm B}) \right] \text{tr}_{\rm A}(\rho_{\rm AB}) \right\} \\
    &= s \cdot \sum_{a,x} \text{tr}_{\rm B} \left\{ F_{a|x} \text{tr}_{\rm A} \left[ (M_{a|x} \otimes \openone_{\rm B})(\xi_{\rm A} \otimes \text{tr}_{\rm A}\rho_{\rm AB}) \right] \right\} \\
    &= s \cdot \text{tr} \sum_{a,x} F_{a|x} \sigma_{a|x}^{\rm m},
\end{aligned}
\end{equation}
where $\sigma_{a|x}^{\rm m} := \text{tr}_{\rm A} \left[ (M_{a|x} \otimes \openone_{\rm B})(\xi_{\rm A} \otimes \text{tr}_{\rm A} \rho_{\rm AB}) \right]$.

Summing both parts, the objective function becomes:
\begin{equation}
    \text{tr}(Z \rho_{\rm AB}) = \text{tr} \sum_{a,x} F_{a|x} \left( \frac{1}{r} \sigma_{a|x} + s \cdot \sigma_{a|x}^{\rm m} \right) = \text{tr} \sum_{a,x} F_{a|x} \rho_{a|x}.
\end{equation}
where $s=(r-1)/r$ mentioned earlier.

Thus, the dual SDP takes the final form:
\begin{equation}
\begin{aligned}
    \min_{F_{a|x}} \quad & \text{tr} \sum_{a,x} F_{a|x} \rho_{a|x} \\
    \text{s.t.} \quad & \sum_{a,x} D(a|x,\lambda) F_{a|x} \succeq 0 \quad \forall \lambda, \\
    & \text{tr} \sum_{a,x,\lambda} D(a|x,\lambda) F_{a|x} = 1
\end{aligned}
\end{equation}

\section{Representation of a steering inequality with local data}\label{Sec_App_Pauli}
In this appendix we show that the objective function of Eq.~\eqref{Eq_dual} can be represented with the probabilities $P(a|x)$ and Bob's local expectation values. A qubit operator can be decomposed with Pauli matrices, namely
\begin{equation}
    F_{a|x} = c_0^{a,x}\openone +c_1^{a,x} X + c_2^{a,x}Y + c_3^{a,x}Z:=\sum_{i} c_i^{a,x} \sigma_i,
\end{equation}
where $\sigma_i$ are the Pauli matrices. Substituting $F_{a|x}$ into the objective function of Eq.~\eqref{Eq_dual}, one then obtains
\begin{equation}
    \begin{aligned}
        \tr\sum_{a,x} F_{a|x}\rho_{a|x} &= \sum_{a,x,i} c_i^{a,x} \tr(\sigma_i\rho_{a|x})\\
        &=\sum_{a,x,i} c_i^{a,x} \big[\frac{1}{r}\tr(\sigma_i\sigma_{a|x}) + \frac{r-1}{r}\tr(\sigma_i\sigma_{a|x}^{\rm m})\big]\\
        &=\sum_{a,x,i}c_i^{a,x}\big[ \frac{1}{r}\cdot P(a|x)\langle\sigma_i\rangle_{\hat{\sigma}_{a|x}} + \frac{r-1}{r}\cdot P_\xi(a|x)\langle\sigma_i\rangle_{\rho_{\rm B}}\big].
    \end{aligned}
\end{equation}
Here, $\langle\sigma_i\rangle_{\hat{\sigma}_{a|x}}$ is the expectation value of Bob's operator $\sigma_i$ when his state conditional on Alice's measurement is $\hat{\sigma}_{a|x}:=\sigma_{a|x}/P(a|x)$, and $\langle\sigma_i\rangle_{\rho_{\rm B}}$ is the expectation value of $\sigma_i$ when Bob receives his half of $\rabout$. Every quantity on the right-hand side is measured on Bob's side alone.

\bibliography{bib_self_testing_states}

\begin{thebibliography}{34}%
\makeatletter
\providecommand \@ifxundefined [1]{%
 \@ifx{#1\undefined}
}%
\providecommand \@ifnum [1]{%
 \ifnum #1\expandafter \@firstoftwo
 \else \expandafter \@secondoftwo
 \fi
}%
\providecommand \@ifx [1]{%
 \ifx #1\expandafter \@firstoftwo
 \else \expandafter \@secondoftwo
 \fi
}%
\providecommand \natexlab [1]{#1}%
\providecommand \enquote  [1]{``#1''}%
\providecommand \bibnamefont  [1]{#1}%
\providecommand \bibfnamefont [1]{#1}%
\providecommand \citenamefont [1]{#1}%
\providecommand \href@noop [0]{\@secondoftwo}%
\providecommand \href [0]{\begingroup \@sanitize@url \@href}%
\providecommand \@href[1]{\@@startlink{#1}\@@href}%
\providecommand \@@href[1]{\endgroup#1\@@endlink}%
\providecommand \@sanitize@url [0]{\catcode `\\12\catcode `\$12\catcode
  `\&12\catcode `\#12\catcode `\^12\catcode `\_12\catcode `\%12\relax}%
\providecommand \@@startlink[1]{}%
\providecommand \@@endlink[0]{}%
\providecommand \url  [0]{\begingroup\@sanitize@url \@url }%
\providecommand \@url [1]{\endgroup\@href {#1}{\urlprefix }}%
\providecommand \urlprefix  [0]{URL }%
\providecommand \Eprint [0]{\href }%
\providecommand \doibase [0]{https://doi.org/}%
\providecommand \selectlanguage [0]{\@gobble}%
\providecommand \bibinfo  [0]{\@secondoftwo}%
\providecommand \bibfield  [0]{\@secondoftwo}%
\providecommand \translation [1]{[#1]}%
\providecommand \BibitemOpen [0]{}%
\providecommand \bibitemStop [0]{}%
\providecommand \bibitemNoStop [0]{.\EOS\space}%
\providecommand \EOS [0]{\spacefactor3000\relax}%
\providecommand \BibitemShut  [1]{\csname bibitem#1\endcsname}%
\let\auto@bib@innerbib\@empty
%</preamble>
\bibitem [{\citenamefont {Cavalcanti}\ and\ \citenamefont
  {Skrzypczyk}(2017)}]{Cavalcanti17}%
  \BibitemOpen
  \bibfield  {author} {\bibinfo {author} {\bibfnamefont {D.}~\bibnamefont
  {Cavalcanti}}\ and\ \bibinfo {author} {\bibfnamefont {P.}~\bibnamefont
  {Skrzypczyk}},\ }\bibfield  {title} {\bibinfo {title} {Quantum steering: a
  review with focus on semidefinite programming},\ }\href
  {http://stacks.iop.org/0034-4885/80/i=2/a=024001} {\bibfield  {journal}
  {\bibinfo  {journal} {Rep. Prof. Phys.}\ }\textbf {\bibinfo {volume} {80}},\
  \bibinfo {pages} {024001} (\bibinfo {year} {2017})}\BibitemShut {NoStop}%
\bibitem [{\citenamefont {Uola}\ \emph {et~al.}(2020)\citenamefont {Uola},
  \citenamefont {Costa}, \citenamefont {Nguyen},\ and\ \citenamefont
  {G\"uhne}}]{Uola2020Steering}%
  \BibitemOpen
  \bibfield  {author} {\bibinfo {author} {\bibfnamefont {R.}~\bibnamefont
  {Uola}}, \bibinfo {author} {\bibfnamefont {A.~C.~S.}\ \bibnamefont {Costa}},
  \bibinfo {author} {\bibfnamefont {H.~C.}\ \bibnamefont {Nguyen}},\ and\
  \bibinfo {author} {\bibfnamefont {O.}~\bibnamefont {G\"uhne}},\ }\bibfield
  {title} {\bibinfo {title} {Quantum steering},\ }\href
  {https://doi.org/10.1103/RevModPhys.92.015001} {\bibfield  {journal}
  {\bibinfo  {journal} {Rev. Mod. Phys.}\ }\textbf {\bibinfo {volume} {92}},\
  \bibinfo {pages} {015001} (\bibinfo {year} {2020})}\BibitemShut {NoStop}%
\bibitem [{\citenamefont {Wiseman}\ \emph {et~al.}(2007)\citenamefont
  {Wiseman}, \citenamefont {Jones},\ and\ \citenamefont {Doherty}}]{Wiseman07}%
  \BibitemOpen
  \bibfield  {author} {\bibinfo {author} {\bibfnamefont {H.~M.}\ \bibnamefont
  {Wiseman}}, \bibinfo {author} {\bibfnamefont {S.~J.}\ \bibnamefont {Jones}},\
  and\ \bibinfo {author} {\bibfnamefont {A.~C.}\ \bibnamefont {Doherty}},\
  }\bibfield  {title} {\bibinfo {title} {Steering, entanglement, nonlocality,
  and the {E}instein-{P}odolsky-{R}osen paradox},\ }\href
  {https://doi.org/10.1103/PhysRevLett.98.140402} {\bibfield  {journal}
  {\bibinfo  {journal} {Phys. Rev. Lett.}\ }\textbf {\bibinfo {volume} {98}},\
  \bibinfo {pages} {140402} (\bibinfo {year} {2007})}\BibitemShut {NoStop}%
\bibitem [{\citenamefont {Pusey}(2013)}]{Pusey13}%
  \BibitemOpen
  \bibfield  {author} {\bibinfo {author} {\bibfnamefont {M.~F.}\ \bibnamefont
  {Pusey}},\ }\bibfield  {title} {\bibinfo {title} {Negativity and steering: A
  stronger peres conjecture},\ }\href
  {https://doi.org/10.1103/PhysRevA.88.032313} {\bibfield  {journal} {\bibinfo
  {journal} {Phys. Rev. A}\ }\textbf {\bibinfo {volume} {88}},\ \bibinfo
  {pages} {032313} (\bibinfo {year} {2013})}\BibitemShut {NoStop}%
\bibitem [{\citenamefont {G\"uhne}\ \emph {et~al.}(2023)\citenamefont
  {G\"uhne}, \citenamefont {Haapasalo}, \citenamefont {Kraft}, \citenamefont
  {Pellonp\"a\"a},\ and\ \citenamefont {Uola}}]{Guhne2023}%
  \BibitemOpen
  \bibfield  {author} {\bibinfo {author} {\bibfnamefont {O.}~\bibnamefont
  {G\"uhne}}, \bibinfo {author} {\bibfnamefont {E.}~\bibnamefont {Haapasalo}},
  \bibinfo {author} {\bibfnamefont {T.}~\bibnamefont {Kraft}}, \bibinfo
  {author} {\bibfnamefont {J.-P.}\ \bibnamefont {Pellonp\"a\"a}},\ and\
  \bibinfo {author} {\bibfnamefont {R.}~\bibnamefont {Uola}},\ }\bibfield
  {title} {\bibinfo {title} {Colloquium: Incompatible measurements in quantum
  information science},\ }\href {https://doi.org/10.1103/RevModPhys.95.011003}
  {\bibfield  {journal} {\bibinfo  {journal} {Rev. Mod. Phys.}\ }\textbf
  {\bibinfo {volume} {95}},\ \bibinfo {pages} {011003} (\bibinfo {year}
  {2023})}\BibitemShut {NoStop}%
\bibitem [{\citenamefont {Lahti}(2003)}]{Lahti03}%
  \BibitemOpen
  \bibfield  {author} {\bibinfo {author} {\bibfnamefont {P.}~\bibnamefont
  {Lahti}},\ }\href {https://doi.org/10.1023/a:1025406103210} {\bibfield
  {journal} {\bibinfo  {journal} {International Journal of Theoretical
  Physics}\ }\textbf {\bibinfo {volume} {42}},\ \bibinfo {pages} {893}
  (\bibinfo {year} {2003})}\BibitemShut {NoStop}%
\bibitem [{\citenamefont {Quintino}\ \emph {et~al.}(2014)\citenamefont
  {Quintino}, \citenamefont {V\'ertesi},\ and\ \citenamefont
  {Brunner}}]{Quint14}%
  \BibitemOpen
  \bibfield  {author} {\bibinfo {author} {\bibfnamefont {M.~T.}\ \bibnamefont
  {Quintino}}, \bibinfo {author} {\bibfnamefont {T.}~\bibnamefont
  {V\'ertesi}},\ and\ \bibinfo {author} {\bibfnamefont {N.}~\bibnamefont
  {Brunner}},\ }\bibfield  {title} {\bibinfo {title} {Joint measurability,
  {E}instein-{P}odolsky-{R}osen steering, and {B}ell nonlocality},\ }\href
  {https://doi.org/10.1103/PhysRevLett.113.160402} {\bibfield  {journal}
  {\bibinfo  {journal} {Phys. Rev. Lett.}\ }\textbf {\bibinfo {volume} {113}},\
  \bibinfo {pages} {160402} (\bibinfo {year} {2014})}\BibitemShut {NoStop}%
\bibitem [{\citenamefont {Uola}\ \emph {et~al.}(2014)\citenamefont {Uola},
  \citenamefont {Moroder},\ and\ \citenamefont {G\"uhne}}]{Uola14}%
  \BibitemOpen
  \bibfield  {author} {\bibinfo {author} {\bibfnamefont {R.}~\bibnamefont
  {Uola}}, \bibinfo {author} {\bibfnamefont {T.}~\bibnamefont {Moroder}},\ and\
  \bibinfo {author} {\bibfnamefont {O.}~\bibnamefont {G\"uhne}},\ }\bibfield
  {title} {\bibinfo {title} {Joint measurability of generalized measurements
  implies classicality},\ }\href
  {https://doi.org/10.1103/PhysRevLett.113.160403} {\bibfield  {journal}
  {\bibinfo  {journal} {Phys. Rev. Lett.}\ }\textbf {\bibinfo {volume} {113}},\
  \bibinfo {pages} {160403} (\bibinfo {year} {2014})}\BibitemShut {NoStop}%
\bibitem [{\citenamefont {Uola}\ \emph {et~al.}(2015)\citenamefont {Uola},
  \citenamefont {Budroni}, \citenamefont {G\"uhne},\ and\ \citenamefont
  {Pellonp\"a\"a}}]{Uola15}%
  \BibitemOpen
  \bibfield  {author} {\bibinfo {author} {\bibfnamefont {R.}~\bibnamefont
  {Uola}}, \bibinfo {author} {\bibfnamefont {C.}~\bibnamefont {Budroni}},
  \bibinfo {author} {\bibfnamefont {O.}~\bibnamefont {G\"uhne}},\ and\ \bibinfo
  {author} {\bibfnamefont {J.-P.}\ \bibnamefont {Pellonp\"a\"a}},\ }\bibfield
  {title} {\bibinfo {title} {One-to-one mapping between steering and joint
  measurability problems},\ }\href
  {https://doi.org/10.1103/PhysRevLett.115.230402} {\bibfield  {journal}
  {\bibinfo  {journal} {Phys. Rev. Lett.}\ }\textbf {\bibinfo {volume} {115}},\
  \bibinfo {pages} {230402} (\bibinfo {year} {2015})}\BibitemShut {NoStop}%
\bibitem [{\citenamefont {Tavakoli}\ and\ \citenamefont
  {Uola}(2020)}]{Tavakoli2020Measurement}%
  \BibitemOpen
  \bibfield  {author} {\bibinfo {author} {\bibfnamefont {A.}~\bibnamefont
  {Tavakoli}}\ and\ \bibinfo {author} {\bibfnamefont {R.}~\bibnamefont
  {Uola}},\ }\bibfield  {title} {\bibinfo {title} {Measurement incompatibility
  and steering are necessary and sufficient for operational contextuality},\
  }\href {https://doi.org/10.1103/PhysRevResearch.2.013011} {\bibfield
  {journal} {\bibinfo  {journal} {Phys. Rev. Research}\ }\textbf {\bibinfo
  {volume} {2}},\ \bibinfo {pages} {013011} (\bibinfo {year}
  {2020})}\BibitemShut {NoStop}%
\bibitem [{\citenamefont {Porto}\ \emph
  {et~al.}(2026{\natexlab{a}})\citenamefont {Porto}, \citenamefont {Tendick},
  \citenamefont {Cavalcanti}, \citenamefont {Uola},\ and\ \citenamefont
  {Quintino}}]{Porto2026Can_arXiv}%
  \BibitemOpen
  \bibfield  {author} {\bibinfo {author} {\bibfnamefont {L.~E.~A.}\
  \bibnamefont {Porto}}, \bibinfo {author} {\bibfnamefont {L.}~\bibnamefont
  {Tendick}}, \bibinfo {author} {\bibfnamefont {D.}~\bibnamefont {Cavalcanti}},
  \bibinfo {author} {\bibfnamefont {R.}~\bibnamefont {Uola}},\ and\ \bibinfo
  {author} {\bibfnamefont {M.~T.}\ \bibnamefont {Quintino}},\ }\href@noop {}
  {\bibinfo {title} {Can every set of incompatible measurements lead to genuine
  multipartite steering?}} (\bibinfo {year} {2026}{\natexlab{a}}),\ \Eprint
  {https://arxiv.org/abs/arXiv:2603.25345} {arXiv:2603.25345} \BibitemShut
  {NoStop}%
\bibitem [{\citenamefont {Porto}\ \emph
  {et~al.}(2026{\natexlab{b}})\citenamefont {Porto}, \citenamefont
  {Designolle}, \citenamefont {Pokutta},\ and\ \citenamefont
  {Quintino}}]{Porto2026Measurement}%
  \BibitemOpen
  \bibfield  {author} {\bibinfo {author} {\bibfnamefont {L.~E.~A.}\
  \bibnamefont {Porto}}, \bibinfo {author} {\bibfnamefont {S.}~\bibnamefont
  {Designolle}}, \bibinfo {author} {\bibfnamefont {S.}~\bibnamefont
  {Pokutta}},\ and\ \bibinfo {author} {\bibfnamefont {M.~T.}\ \bibnamefont
  {Quintino}},\ }\bibfield  {title} {\bibinfo {title} {Measurement
  incompatibility and quantum steering via linear programming},\ }\href
  {https://doi.org/10.22331/q-2026-06-19-2141} {\bibfield  {journal} {\bibinfo
  {journal} {Quantum}\ }\textbf {\bibinfo {volume} {10}},\ \bibinfo {pages}
  {2141} (\bibinfo {year} {2026}{\natexlab{b}})}\BibitemShut {NoStop}%
\bibitem [{\citenamefont {Wolf}\ \emph {et~al.}(2009)\citenamefont {Wolf},
  \citenamefont {Perez-Garcia},\ and\ \citenamefont {Fernandez}}]{Wolf09}%
  \BibitemOpen
  \bibfield  {author} {\bibinfo {author} {\bibfnamefont {M.~M.}\ \bibnamefont
  {Wolf}}, \bibinfo {author} {\bibfnamefont {D.}~\bibnamefont {Perez-Garcia}},\
  and\ \bibinfo {author} {\bibfnamefont {C.}~\bibnamefont {Fernandez}},\
  }\bibfield  {title} {\bibinfo {title} {Measurements incompatible in quantum
  theory cannot be measured jointly in any other no-signaling theory},\ }\href
  {https://doi.org/10.1103/PhysRevLett.103.230402} {\bibfield  {journal}
  {\bibinfo  {journal} {Phys. Rev. Lett.}\ }\textbf {\bibinfo {volume} {103}},\
  \bibinfo {pages} {230402} (\bibinfo {year} {2009})}\BibitemShut {NoStop}%
\bibitem [{\citenamefont {Xu}\ and\ \citenamefont {Cabello}(2019)}]{Xu19}%
  \BibitemOpen
  \bibfield  {author} {\bibinfo {author} {\bibfnamefont {Z.-P.}\ \bibnamefont
  {Xu}}\ and\ \bibinfo {author} {\bibfnamefont {A.}~\bibnamefont {Cabello}},\
  }\bibfield  {title} {\bibinfo {title} {Necessary and sufficient condition for
  contextuality from incompatibility},\ }\href
  {https://doi.org/10.1103/PhysRevA.99.020103} {\bibfield  {journal} {\bibinfo
  {journal} {Phys. Rev. A}\ }\textbf {\bibinfo {volume} {99}},\ \bibinfo
  {pages} {020103} (\bibinfo {year} {2019})}\BibitemShut {NoStop}%
\bibitem [{\citenamefont {Ku}\ \emph {et~al.}(2022)\citenamefont {Ku},
  \citenamefont {Kadlec}, \citenamefont {\ifmmode~\check{C}\else
  \v{C}\fi{}ernoch}, \citenamefont {Quintino}, \citenamefont {Zhou},
  \citenamefont {Lemr}, \citenamefont {Lambert}, \citenamefont {Miranowicz},
  \citenamefont {Chen}, \citenamefont {Nori},\ and\ \citenamefont
  {Chen}}]{HYKu2022}%
  \BibitemOpen
  \bibfield  {author} {\bibinfo {author} {\bibfnamefont {H.-Y.}\ \bibnamefont
  {Ku}}, \bibinfo {author} {\bibfnamefont {J.}~\bibnamefont {Kadlec}}, \bibinfo
  {author} {\bibfnamefont {A.}~\bibnamefont {\ifmmode~\check{C}\else
  \v{C}\fi{}ernoch}}, \bibinfo {author} {\bibfnamefont {M.~T.}\ \bibnamefont
  {Quintino}}, \bibinfo {author} {\bibfnamefont {W.}~\bibnamefont {Zhou}},
  \bibinfo {author} {\bibfnamefont {K.}~\bibnamefont {Lemr}}, \bibinfo {author}
  {\bibfnamefont {N.}~\bibnamefont {Lambert}}, \bibinfo {author} {\bibfnamefont
  {A.}~\bibnamefont {Miranowicz}}, \bibinfo {author} {\bibfnamefont {S.-L.}\
  \bibnamefont {Chen}}, \bibinfo {author} {\bibfnamefont {F.}~\bibnamefont
  {Nori}},\ and\ \bibinfo {author} {\bibfnamefont {Y.-N.}\ \bibnamefont
  {Chen}},\ }\bibfield  {title} {\bibinfo {title} {Quantifying quantumness of
  channels without entanglement},\ }\href
  {https://doi.org/10.1103/PRXQuantum.3.020338} {\bibfield  {journal} {\bibinfo
   {journal} {PRX Quantum}\ }\textbf {\bibinfo {volume} {3}},\ \bibinfo {pages}
  {020338} (\bibinfo {year} {2022})}\BibitemShut {NoStop}%
\bibitem [{\citenamefont {Heinosaari}\ \emph {et~al.}(2015)\citenamefont
  {Heinosaari}, \citenamefont {Kiukas}, \citenamefont {Reitzner},\ and\
  \citenamefont {Schultz}}]{Heinosaari2015}%
  \BibitemOpen
  \bibfield  {author} {\bibinfo {author} {\bibfnamefont {T.}~\bibnamefont
  {Heinosaari}}, \bibinfo {author} {\bibfnamefont {J.}~\bibnamefont {Kiukas}},
  \bibinfo {author} {\bibfnamefont {D.}~\bibnamefont {Reitzner}},\ and\
  \bibinfo {author} {\bibfnamefont {J.}~\bibnamefont {Schultz}},\ }\bibfield
  {title} {\bibinfo {title} {Incompatibility breaking quantum channels},\
  }\href {https://doi.org/10.1088/1751-8113/48/43/435301} {\bibfield  {journal}
  {\bibinfo  {journal} {Journal of Physics A: Mathematical and Theoretical}\
  }\textbf {\bibinfo {volume} {48}},\ \bibinfo {pages} {435301} (\bibinfo
  {year} {2015})}\BibitemShut {NoStop}%
\bibitem [{\citenamefont {Horodecki}\ \emph {et~al.}(2003)\citenamefont
  {Horodecki}, \citenamefont {Shor},\ and\ \citenamefont
  {Ruskai}}]{Horodecki2003}%
  \BibitemOpen
  \bibfield  {author} {\bibinfo {author} {\bibfnamefont {M.}~\bibnamefont
  {Horodecki}}, \bibinfo {author} {\bibfnamefont {P.~W.}\ \bibnamefont
  {Shor}},\ and\ \bibinfo {author} {\bibfnamefont {M.~B.}\ \bibnamefont
  {Ruskai}},\ }\bibfield  {title} {\bibinfo {title} {Entanglement breaking
  channels},\ }\href {https://doi.org/10.1142/s0129055x03001709} {\bibfield
  {journal} {\bibinfo  {journal} {Reviews in Mathematical Physics}\ }\textbf
  {\bibinfo {volume} {15}},\ \bibinfo {pages} {629–641} (\bibinfo {year}
  {2003})}\BibitemShut {NoStop}%
\bibitem [{\citenamefont {Ruskai}(2003)}]{Ruskai2003}%
  \BibitemOpen
  \bibfield  {author} {\bibinfo {author} {\bibfnamefont {M.~B.}\ \bibnamefont
  {Ruskai}},\ }\bibfield  {title} {\bibinfo {title} {Qubit entanglement
  breaking channels},\ }\href {https://doi.org/10.1142/s0129055x03001710}
  {\bibfield  {journal} {\bibinfo  {journal} {Reviews in Mathematical Physics}\
  }\textbf {\bibinfo {volume} {15}},\ \bibinfo {pages} {643–662} (\bibinfo
  {year} {2003})}\BibitemShut {NoStop}%
\bibitem [{\citenamefont {Cavalcanti}\ \emph {et~al.}(2016)\citenamefont
  {Cavalcanti}, \citenamefont {Guerini}, \citenamefont {Rabelo},\ and\
  \citenamefont {Skrzypczyk}}]{Cavalcanti2016}%
  \BibitemOpen
  \bibfield  {author} {\bibinfo {author} {\bibfnamefont {D.}~\bibnamefont
  {Cavalcanti}}, \bibinfo {author} {\bibfnamefont {L.}~\bibnamefont {Guerini}},
  \bibinfo {author} {\bibfnamefont {R.}~\bibnamefont {Rabelo}},\ and\ \bibinfo
  {author} {\bibfnamefont {P.}~\bibnamefont {Skrzypczyk}},\ }\bibfield  {title}
  {\bibinfo {title} {General method for constructing local hidden variable
  models for entangled quantum states},\ }\href
  {https://doi.org/10.1103/PhysRevLett.117.190401} {\bibfield  {journal}
  {\bibinfo  {journal} {Phys. Rev. Lett.}\ }\textbf {\bibinfo {volume} {117}},\
  \bibinfo {pages} {190401} (\bibinfo {year} {2016})}\BibitemShut {NoStop}%
\bibitem [{\citenamefont {Hirsch}\ \emph {et~al.}(2016)\citenamefont {Hirsch},
  \citenamefont {Quintino}, \citenamefont {V\'ertesi}, \citenamefont {Pusey},\
  and\ \citenamefont {Brunner}}]{Hirsch2016}%
  \BibitemOpen
  \bibfield  {author} {\bibinfo {author} {\bibfnamefont {F.}~\bibnamefont
  {Hirsch}}, \bibinfo {author} {\bibfnamefont {M.~T.}\ \bibnamefont
  {Quintino}}, \bibinfo {author} {\bibfnamefont {T.}~\bibnamefont {V\'ertesi}},
  \bibinfo {author} {\bibfnamefont {M.~F.}\ \bibnamefont {Pusey}},\ and\
  \bibinfo {author} {\bibfnamefont {N.}~\bibnamefont {Brunner}},\ }\bibfield
  {title} {\bibinfo {title} {Algorithmic construction of local hidden variable
  models for entangled quantum states},\ }\href
  {https://doi.org/10.1103/PhysRevLett.117.190402} {\bibfield  {journal}
  {\bibinfo  {journal} {Phys. Rev. Lett.}\ }\textbf {\bibinfo {volume} {117}},\
  \bibinfo {pages} {190402} (\bibinfo {year} {2016})}\BibitemShut {NoStop}%
\bibitem [{\citenamefont {Skrzypczyk}\ \emph {et~al.}(2014)\citenamefont
  {Skrzypczyk}, \citenamefont {Navascu\'es},\ and\ \citenamefont
  {Cavalcanti}}]{SNC14}%
  \BibitemOpen
  \bibfield  {author} {\bibinfo {author} {\bibfnamefont {P.}~\bibnamefont
  {Skrzypczyk}}, \bibinfo {author} {\bibfnamefont {M.}~\bibnamefont
  {Navascu\'es}},\ and\ \bibinfo {author} {\bibfnamefont {D.}~\bibnamefont
  {Cavalcanti}},\ }\bibfield  {title} {\bibinfo {title} {Quantifying
  {E}instein-{P}odolsky-{R}osen steering},\ }\href
  {https://doi.org/10.1103/PhysRevLett.112.180404} {\bibfield  {journal}
  {\bibinfo  {journal} {Phys. Rev. Lett.}\ }\textbf {\bibinfo {volume} {112}},\
  \bibinfo {pages} {180404} (\bibinfo {year} {2014})}\BibitemShut {NoStop}%
\bibitem [{\citenamefont {Werner}(1989)}]{Werner89}%
  \BibitemOpen
  \bibfield  {author} {\bibinfo {author} {\bibfnamefont {R.~F.}\ \bibnamefont
  {Werner}},\ }\bibfield  {title} {\bibinfo {title} {Quantum states with
  {E}instein-{P}odolsky-{R}osen correlations admitting a hidden-variable
  model},\ }\href {https://doi.org/10.1103/PhysRevA.40.4277} {\bibfield
  {journal} {\bibinfo  {journal} {Phys. Rev. A}\ }\textbf {\bibinfo {volume}
  {40}},\ \bibinfo {pages} {4277} (\bibinfo {year} {1989})}\BibitemShut
  {NoStop}%
\bibitem [{\citenamefont {Bowles}\ \emph {et~al.}(2014)\citenamefont {Bowles},
  \citenamefont {V\'ertesi}, \citenamefont {Quintino},\ and\ \citenamefont
  {Brunner}}]{Bowles2014}%
  \BibitemOpen
  \bibfield  {author} {\bibinfo {author} {\bibfnamefont {J.}~\bibnamefont
  {Bowles}}, \bibinfo {author} {\bibfnamefont {T.}~\bibnamefont {V\'ertesi}},
  \bibinfo {author} {\bibfnamefont {M.~T.}\ \bibnamefont {Quintino}},\ and\
  \bibinfo {author} {\bibfnamefont {N.}~\bibnamefont {Brunner}},\ }\bibfield
  {title} {\bibinfo {title} {One-way {E}instein-{P}odolsky-{R}osen steering},\
  }\href {https://doi.org/10.1103/PhysRevLett.112.200402} {\bibfield  {journal}
  {\bibinfo  {journal} {Phys. Rev. Lett.}\ }\textbf {\bibinfo {volume} {112}},\
  \bibinfo {pages} {200402} (\bibinfo {year} {2014})}\BibitemShut {NoStop}%
\bibitem [{\citenamefont {Jevtic}\ \emph {et~al.}(2015)\citenamefont {Jevtic},
  \citenamefont {Hall}, \citenamefont {Anderson}, \citenamefont {Zwierz},\ and\
  \citenamefont {Wiseman}}]{Jevtic2015}%
  \BibitemOpen
  \bibfield  {author} {\bibinfo {author} {\bibfnamefont {S.}~\bibnamefont
  {Jevtic}}, \bibinfo {author} {\bibfnamefont {M.~J.~W.}\ \bibnamefont {Hall}},
  \bibinfo {author} {\bibfnamefont {M.~R.}\ \bibnamefont {Anderson}}, \bibinfo
  {author} {\bibfnamefont {M.}~\bibnamefont {Zwierz}},\ and\ \bibinfo {author}
  {\bibfnamefont {H.~M.}\ \bibnamefont {Wiseman}},\ }\bibfield  {title}
  {\bibinfo {title} {{E}instein–{P}odolsky–{R}osen steering and the
  steering ellipsoid},\ }\href {https://doi.org/10.1364/josab.32.000a40}
  {\bibfield  {journal} {\bibinfo  {journal} {Journal of the Optical Society of
  America B}\ }\textbf {\bibinfo {volume} {32}},\ \bibinfo {pages} {A40}
  (\bibinfo {year} {2015})}\BibitemShut {NoStop}%
\bibitem [{\citenamefont {Bowles}\ \emph {et~al.}(2015)\citenamefont {Bowles},
  \citenamefont {Hirsch}, \citenamefont {Quintino},\ and\ \citenamefont
  {Brunner}}]{Bowles2015}%
  \BibitemOpen
  \bibfield  {author} {\bibinfo {author} {\bibfnamefont {J.}~\bibnamefont
  {Bowles}}, \bibinfo {author} {\bibfnamefont {F.}~\bibnamefont {Hirsch}},
  \bibinfo {author} {\bibfnamefont {M.~T.}\ \bibnamefont {Quintino}},\ and\
  \bibinfo {author} {\bibfnamefont {N.}~\bibnamefont {Brunner}},\ }\bibfield
  {title} {\bibinfo {title} {Local hidden variable models for entangled quantum
  states using finite shared randomness},\ }\href
  {https://doi.org/10.1103/PhysRevLett.114.120401} {\bibfield  {journal}
  {\bibinfo  {journal} {Phys. Rev. Lett.}\ }\textbf {\bibinfo {volume} {114}},\
  \bibinfo {pages} {120401} (\bibinfo {year} {2015})}\BibitemShut {NoStop}%
\bibitem [{\citenamefont {Hirsch}\ \emph {et~al.}(2017)\citenamefont {Hirsch},
  \citenamefont {Quintino}, \citenamefont {Vértesi}, \citenamefont
  {Navascués},\ and\ \citenamefont {Brunner}}]{Hirsch2017}%
  \BibitemOpen
  \bibfield  {author} {\bibinfo {author} {\bibfnamefont {F.}~\bibnamefont
  {Hirsch}}, \bibinfo {author} {\bibfnamefont {M.~T.}\ \bibnamefont
  {Quintino}}, \bibinfo {author} {\bibfnamefont {T.}~\bibnamefont {Vértesi}},
  \bibinfo {author} {\bibfnamefont {M.}~\bibnamefont {Navascués}},\ and\
  \bibinfo {author} {\bibfnamefont {N.}~\bibnamefont {Brunner}},\ }\bibfield
  {title} {\bibinfo {title} {Better local hidden variable models for two-qubit
  {W}erner states and an upper bound on the {G}rothendieck constant
  ${K}_{G}(3)$},\ }\href {https://doi.org/10.22331/q-2017-04-25-3} {\bibfield
  {journal} {\bibinfo  {journal} {Quantum}\ }\textbf {\bibinfo {volume} {1}},\
  \bibinfo {pages} {3} (\bibinfo {year} {2017})}\BibitemShut {NoStop}%
\bibitem [{\citenamefont {Zhang}\ and\ \citenamefont
  {Chitambar}(2024)}]{ZhangYJ2024}%
  \BibitemOpen
  \bibfield  {author} {\bibinfo {author} {\bibfnamefont {Y.}~\bibnamefont
  {Zhang}}\ and\ \bibinfo {author} {\bibfnamefont {E.}~\bibnamefont
  {Chitambar}},\ }\bibfield  {title} {\bibinfo {title} {Exact steering bound
  for two-qubit {W}erner states},\ }\href
  {https://doi.org/10.1103/PhysRevLett.132.250201} {\bibfield  {journal}
  {\bibinfo  {journal} {Phys. Rev. Lett.}\ }\textbf {\bibinfo {volume} {132}},\
  \bibinfo {pages} {250201} (\bibinfo {year} {2024})}\BibitemShut {NoStop}%
\bibitem [{\citenamefont {Ac\'{\i}n}\ \emph {et~al.}(2006)\citenamefont
  {Ac\'{\i}n}, \citenamefont {Gisin},\ and\ \citenamefont {Toner}}]{Acin2006}%
  \BibitemOpen
  \bibfield  {author} {\bibinfo {author} {\bibfnamefont {A.}~\bibnamefont
  {Ac\'{\i}n}}, \bibinfo {author} {\bibfnamefont {N.}~\bibnamefont {Gisin}},\
  and\ \bibinfo {author} {\bibfnamefont {B.}~\bibnamefont {Toner}},\ }\bibfield
   {title} {\bibinfo {title} {Grothendieck's constant and local models for
  noisy entangled quantum states},\ }\href
  {https://doi.org/10.1103/PhysRevA.73.062105} {\bibfield  {journal} {\bibinfo
  {journal} {Phys. Rev. A}\ }\textbf {\bibinfo {volume} {73}},\ \bibinfo
  {pages} {062105} (\bibinfo {year} {2006})}\BibitemShut {NoStop}%
\bibitem [{\citenamefont {Carmeli}\ \emph {et~al.}(2019)\citenamefont
  {Carmeli}, \citenamefont {Heinosaari},\ and\ \citenamefont
  {Toigo}}]{Carmeli19}%
  \BibitemOpen
  \bibfield  {author} {\bibinfo {author} {\bibfnamefont {C.}~\bibnamefont
  {Carmeli}}, \bibinfo {author} {\bibfnamefont {T.}~\bibnamefont
  {Heinosaari}},\ and\ \bibinfo {author} {\bibfnamefont {A.}~\bibnamefont
  {Toigo}},\ }\bibfield  {title} {\bibinfo {title} {Quantum incompatibility
  witnesses},\ }\href {https://doi.org/10.1103/PhysRevLett.122.130402}
  {\bibfield  {journal} {\bibinfo  {journal} {Phys. Rev. Lett.}\ }\textbf
  {\bibinfo {volume} {122}},\ \bibinfo {pages} {130402} (\bibinfo {year}
  {2019})}\BibitemShut {NoStop}%
\bibitem [{\citenamefont {Skrzypczyk}\ \emph {et~al.}(2019)\citenamefont
  {Skrzypczyk}, \citenamefont {\ifmmode \check{S}\else
  \v{S}\fi{}upi\ifmmode~\acute{c}\else \'{c}\fi{}},\ and\ \citenamefont
  {Cavalcanti}}]{Skrzypczyk19}%
  \BibitemOpen
  \bibfield  {author} {\bibinfo {author} {\bibfnamefont {P.}~\bibnamefont
  {Skrzypczyk}}, \bibinfo {author} {\bibfnamefont {I.}~\bibnamefont {\ifmmode
  \check{S}\else \v{S}\fi{}upi\ifmmode~\acute{c}\else \'{c}\fi{}}},\ and\
  \bibinfo {author} {\bibfnamefont {D.}~\bibnamefont {Cavalcanti}},\ }\bibfield
   {title} {\bibinfo {title} {All sets of incompatible measurements give an
  advantage in quantum state discrimination},\ }\href
  {https://doi.org/10.1103/PhysRevLett.122.130403} {\bibfield  {journal}
  {\bibinfo  {journal} {Phys. Rev. Lett.}\ }\textbf {\bibinfo {volume} {122}},\
  \bibinfo {pages} {130403} (\bibinfo {year} {2019})}\BibitemShut {NoStop}%
\bibitem [{\citenamefont {Uola}\ \emph {et~al.}(2019)\citenamefont {Uola},
  \citenamefont {Kraft}, \citenamefont {Shang}, \citenamefont {Yu},\ and\
  \citenamefont {G\"uhne}}]{Uola19a}%
  \BibitemOpen
  \bibfield  {author} {\bibinfo {author} {\bibfnamefont {R.}~\bibnamefont
  {Uola}}, \bibinfo {author} {\bibfnamefont {T.}~\bibnamefont {Kraft}},
  \bibinfo {author} {\bibfnamefont {J.}~\bibnamefont {Shang}}, \bibinfo
  {author} {\bibfnamefont {X.-D.}\ \bibnamefont {Yu}},\ and\ \bibinfo {author}
  {\bibfnamefont {O.}~\bibnamefont {G\"uhne}},\ }\bibfield  {title} {\bibinfo
  {title} {Quantifying quantum resources with conic programming},\ }\href
  {https://doi.org/10.1103/PhysRevLett.122.130404} {\bibfield  {journal}
  {\bibinfo  {journal} {Phys. Rev. Lett.}\ }\textbf {\bibinfo {volume} {122}},\
  \bibinfo {pages} {130404} (\bibinfo {year} {2019})}\BibitemShut {NoStop}%
\bibitem [{\citenamefont {Designolle}\ \emph {et~al.}(2019)\citenamefont
  {Designolle}, \citenamefont {Farkas},\ and\ \citenamefont
  {Kaniewski}}]{Designolle19}%
  \BibitemOpen
  \bibfield  {author} {\bibinfo {author} {\bibfnamefont {S.}~\bibnamefont
  {Designolle}}, \bibinfo {author} {\bibfnamefont {M.}~\bibnamefont {Farkas}},\
  and\ \bibinfo {author} {\bibfnamefont {J.}~\bibnamefont {Kaniewski}},\
  }\bibfield  {title} {\bibinfo {title} {Incompatibility robustness of quantum
  measurements: a unified framework},\ }\href
  {https://doi.org/10.1088/1367-2630/ab5020} {\bibfield  {journal} {\bibinfo
  {journal} {New J. Phys.}\ }\textbf {\bibinfo {volume} {21}},\ \bibinfo
  {pages} {113053} (\bibinfo {year} {2019})}\BibitemShut {NoStop}%
\bibitem [{PTH()}]{PTHsu2026_code}%
  \BibitemOpen
  \href@noop {} {}\bibinfo {note} {Available at
  \url{https://github.com/PoTingHsu/steering-breaking-channels}}\BibitemShut
  {NoStop}%
\bibitem [{\citenamefont {Zhang}(2026)}]{YJZhang2026Exact}%
  \BibitemOpen
  \bibfield  {author} {\bibinfo {author} {\bibfnamefont {Y.}~\bibnamefont
  {Zhang}},\ }\href@noop {} {\bibinfo {title} {Exact incompatibility-breaking
  criterion for unital qubit channels}} (\bibinfo {year} {2026}),\ \Eprint
  {https://arxiv.org/abs/arXiv:2607.27757} {arXiv:2607.27757} \BibitemShut
  {NoStop}%
\end{thebibliography}%

\clearpage
\onecolumngrid

\end{document}